\documentclass[letterpaper,aps,prl,superscriptaddress,showpacs,floatfix,twocolumn]{revtex4}

\usepackage{graphicx}
\usepackage{hyperref}
\usepackage{amsmath}
\usepackage{url}
\usepackage{colordvi}
\usepackage{subfigure}

\usepackage{times}

\begin{document}

\title{Dependencies of lepton angular 
distribution coefficients on the transverse momentum and rapidity of
$Z$ bosons produced in $pp$ collisions at the LHC}

\author{Wen-Chen Chang}
\affiliation{Institute of Physics, Academia Sinica, Taipei 11529, Taiwan}

\author{Randall Evan McClellan}
\affiliation{Department of Physics, University of Illinois at
Urbana-Champaign, Urbana, Illinois 61801, USA}
\affiliation{Thomas Jefferson National Accelerator Facility,
Newport News, Virginia 23606, USA}

\author{Jen-Chieh Peng}
\affiliation{Department of Physics, University of Illinois at
Urbana-Champaign, Urbana, Illinois 61801, USA}

\author{Oleg Teryaev}
\affiliation{Bogoliubov Laboratory of Theoretical Physics,
JINR, 141980 Dubna, Russia}

\date{\today}

\begin{abstract}
High precision data of lepton angular distributions for $\gamma^*/Z$ 
production in $pp$ collisions at the LHC, covering broad ranges of 
dilepton 
transverse momenta ($q_T$) and rapidity ($y$), were recently reported. 
Strong $q_T$ dependencies were observed for several angular distribution
coefficients, $A_i$, including $A_0 - A_4$. Significant 
$y$ dependencies were also found
for the coefficients $A_1$, $A_3$ and $A_4$, while $A_0$ and $A_2$
exhibit very weak rapidity dependence. Using an intuitive geometric
picture, we show that the $q_T$ and $y$ dependencies of 
the angular distributions coefficients can be well described. 
\end{abstract}

\maketitle

\section {I. Introduction}

The angular distribution of leptons produced in the Drell-Yan
process~\cite{drell} remains a subject of considerable interest. 
The original Drell-Yan model offered a specific prediction of
a transversely polarized virtual photon for collinear
quark-antiquark annihilation, resulting in a $1+\cos^2 \theta$
lepton angular distribution~\cite{drell}. This prediction was
in good agreement with the earliest data, which were dominantly
from dileptons with low transverse momentum ($q_T$)~\cite{kenyon,plm}.
As the dilepton's
transverse momentum becomes large, due to
QCD effects involving emission of partons of large
transverse momenta, the 
angular distribution would no longer be azimuthally symmetric.
A general expression for the lepton
angular distribution in the Drell-Yan process becomes~\cite{lam78}

\begin{eqnarray}
\frac{d\sigma}{d\Omega} \propto 1+ \lambda \cos^2\theta 
+ \mu \sin 2 \theta\cos\phi 
+ \frac{\nu}{2} \sin^2\theta \cos 2 \phi,
\label{eq:eq1}
\end{eqnarray}
where $\theta$ and $\phi$ refer to the polar and azimuthal angles of $l^-$
($e^-$ or $\mu^-$) in the rest frame of $\gamma^*$.
The azimuthal dependencies of the
lepton angular distributions are described by the parameters $\mu$ and $\nu$.
While $\lambda = 1, \mu=0,$ and $\nu=0$ in the original 
Drell-Yan model~\cite{drell}, the
presence of the intrinsic transverse momentum and QCD effects
would allow $\lambda \ne 1$ and $\mu, \nu \ne 0$. However,
it was predicted~\cite{lam78} that the deviation
of $\lambda$ from unity is precisely correlated with the coefficient 
of the
$\cos 2 \phi$ term, namely,
$1-\lambda = 2 \nu$. This so-called Lam-Tung relation,
expected to be insensitive to QCD 
corrections~\cite{collins,lam80,boer,berger07}, was
found to be significantly violated in pion-induced Drell-Yan
experiments~\cite{falciano86,conway}.
The unexpectedly large violation of the Lam-Tung relation  
inspired many theoretical
work~\cite{brandenburg94,eskola94,brandenburg93,boer99}, 
including the suggestion~\cite{boer99} that a nonperturbative effect
originating from the novel transverse-momentum-dependent (TMD) Boer-Mulders 
function~\cite{boer98} can account for this violation. This suggestion 
was found to be consistent with the existing pion and proton induced Drell-Yan 
data~\cite{zhu}. It also led to first extractions of the
Boer-Mulders functions from the $\cos 2 \phi$ 
dependence of the unpolarized Drell-Yan data~\cite{bqma,barone10}.
The azimuthal angular distributions
of leptons in unpolarized or polarized Drell-Yan process 
are now regarded as an important tool for accessing the 
novel TMDs~\cite{boer99,metz,jin,peng_qiu}. 

At collider energies, measurement of lepton angular distributions
in $W$ and $Z$ boson productions has long been advocated as 
a sensitive tool for understanding the production mechanism of these gauge
bosons~\cite{mirkes,berger}. The first measurement of the lepton angular
distribution in $\gamma^*/Z$ production was reported by the CDF Collaboration
for $\bar p p$ collision at 1.96 TeV~\cite{CDF}. 
Very recently, the CMS~\cite{cms} and 
ATLAS~\cite{atlas} Collaborations at the LHC reported high-statistics 
measurements of the 
lepton angular distribution of $\gamma^*/Z$ production in $p p$ collision
at $\sqrt s = 8$ TeV. 
Strong $q_T$ dependencies were observed for the $\lambda, \mu$, and $\nu$
parameters. Moreover, violation of the Lam-Tung relation was 
found for these data at large $q_T$. Since the effects of TMD are
expected to be negligible at large $q_T$, the presence of the 
Boer-Mulders function
cannot explain the striking violation of the Lam-Tung relation
at LHC energies.

In a recent paper~\cite{peng16}, we showed that the observed $q_T$
dependence of $\lambda$ and $\nu$, as well as the 
violation of the Lam-Tung relation, can be well described by
a geometric picture. While it is important to compare
perturbative QCD calculations with these data, it is also
instructive to understand the essential features of these data
in terms of an intuitive geometric picture.
In this paper, we extend the previous
work, which focuses on the $\lambda$ and $\nu$ parameters
and the Lam-Tung relation, to other angular distribution parameters.
We also compare the striking $q_T$ and rapidity ($y$) dependencies of 
the angular distribution coefficients measured at the LHC with
our intuitive geometric picture. We find that many salient features
of the data can be well understood within the framework of this simple
and intuitive approach. 

This paper is organized as follows. In Sec. II we present
our model and derive some expressions relevant for understanding
the lepton angular distributions for $\gamma^*/Z$ production. We 
then compare 
calculations using this model with data on the $q_T$ and rapidity
dependencies in Secs. III and IV, respectively. 
We conclude in Sec. V. 

\section {II. Lepton Angular Distribution Coefficients}

The lepton angular distribution in the $\gamma^*/Z$ rest frame is expressed
by both the CMS and ATLAS Collaborations as
\begin{eqnarray}
\frac{d\sigma}{d\Omega} & \propto & (1+\cos^2\theta)+\frac{A_0}{2}
(1-3\cos^2\theta)+A_1 \sin 2 \theta\cos\phi \nonumber \\
& + & \frac{A_2}{2} \sin^2\theta \cos 2 \phi
+ A_3 \sin\theta \cos\phi + A_4 \cos\theta \nonumber \\
& + & A_5 \sin^2\theta \sin 2\phi
+ A_6 \sin 2\theta \sin\phi \nonumber \\
& + & A_7 \sin\theta \sin\phi,
\label{eq:eq2}
\end{eqnarray}
where $\theta$ and $\phi$ are the polar and azimuthal angles of $l^-$
($e^-$ or $\mu^-$) in the rest frame of $\gamma^*/Z$ like 
in Eq.~(\ref{eq:eq1}). Compared to Eq.~(\ref{eq:eq1}),
Eq.~(\ref{eq:eq2}) contains several additional
terms ($A_3 - A_7$), due to the presence of parity-violating
coupling for the $Z$ boson. It is clear
that $\lambda, \mu, \nu$ in Eq. (1) are related to $A_0, A_1, A_2$
via
\begin{eqnarray}
\lambda = \frac{2-3A_0}{2+A_0};~~~ \mu  =  \frac{2A_1}{2+A_0};~~~
\nu  =  \frac{2A_2}{2+A_0}.
\label{eq:eq3}
\end{eqnarray}
Equation (\ref{eq:eq3}) shows that the Lam-Tung relation,
$1-\lambda = 2 \nu$, becomes $A_0 = A_2$.

\begin{figure}[tb]
\includegraphics*[width=\linewidth]{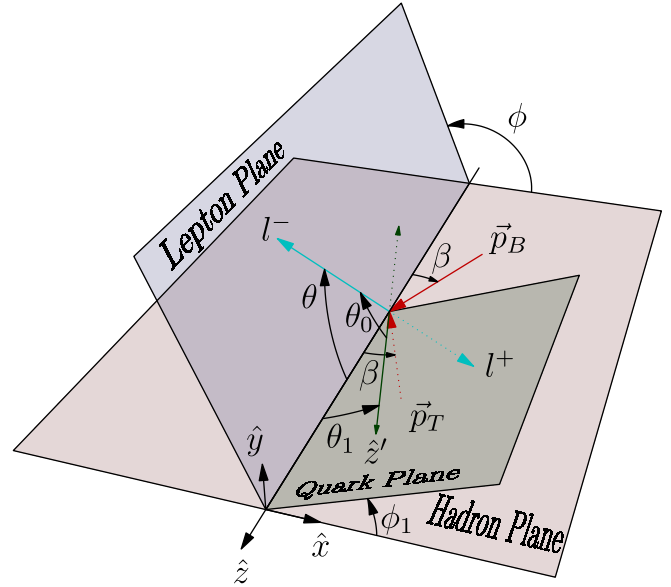}
\caption{Definition of the Collins-Soper frame and various angles
and planes in the rest frame of $\gamma^*/Z$. The hadron plane
is formed by $\vec P_B$ and $\vec P_T$, the momentum vectors of
the beam (B) and target (T) hadrons. The $\hat x$ and $\hat z$ axes
of the Collins-Soper frame both lie in the hadron plane with the
$\hat z$ axis bisecting the $\vec P_B$ and $- \vec P_T$ vectors.
The quark ($q$) and antiquark ($\bar q$) annihilate collinearly
with equal momenta to form $\gamma^*/Z$, while the quark momentum
vector $\hat z^\prime$ and the $\hat z$ axis form the quark plane.
The polar and azimuthal angles of $\hat z^\prime$ in the Collins-Soper
frame are $\theta_1$ and $\phi_1$. The $l^-$ and $l^+$ are emitted
back-to-back with $\theta$ and $\phi$ as the polar and azimuthal angles
for $l^-$.}
\label{fig1}
\end{figure}

While Eq.~(\ref{eq:eq2}) can be derived from the consideration of
the general form of the lepton and hadron tensors involved in 
the $\gamma^*/Z$ production, we present a derivation based on an 
intuitive geometric picture. 
We first define three different planes, i.e., the
hadron plane, the quark plane, and the lepton plane,
shown in Fig.~\ref{fig1}.
For nonzero $q_T$, the beam and target hadron momenta,
$\vec P_B$ and $\vec P_T$, are
no longer collinear in the rest frame of $\gamma^*/Z$, and they
form the ``hadron plane" shown in Fig.~\ref{fig1}. 
Various coordinate systems in the $\gamma^*/Z$ rest
frame have been considered in the literature, and 
the Collins-Soper (C-S) frame~\cite{cs} was used by both the 
CMS and ATLAS
Collaborations. For the Collins-Soper 
frame, the $\hat x$ and $\hat z$ axes both lie in the hadron plane, while the
$\hat z$ axis bisects $\vec P_B$ and $- \vec P_T$ with an angle $\beta$. 
It is straightforward to show that 
\begin{equation}
\tan \beta = q_T / Q,
\label{eq:eq4}
\end{equation}
where $Q$ is the mass of the dilepton. Figure~\ref{fig1}
also shows the ``lepton plane" formed by the 
momentum vector of $l^-$ and the
$\hat z$ axis. The $l^-$ and $l^+$ are emitted back-to-back 
with equal momenta in the rest frame of $\gamma^*/Z$.

In the $\gamma^*/Z$ rest frame, a pair of collinear $q$ and $\bar q$ 
with equal
momenta annihilate into a $\gamma^*/Z$, as illustrated in Fig.~\ref{fig1}.
We define the momentum unit vector of $q$ as $\hat z^\prime$, and the
``quark plane" is formed by the $\hat z^\prime$ and $\hat z$ axes.
The polar and azimuthal angles of the $\hat z^\prime$ axis in 
the Collins-Soper frame are denoted as $\theta_1$ 
and $\phi_1$. The $q -\bar q$
axis, called the ``natural" axis, has the important property~\cite{faccioli}
that the $l^-$ angular distribution is azimuthally symmetric with respect
to this axis, namely,
\begin{equation}
\frac{d\sigma}{d\Omega} \propto  1 + a \cos \theta_0 + \cos^2\theta_0,
\label{eq:eq5}
\end{equation}
where $\theta_0$ is the angle between the $l^-$
momentum vector and the $\hat z^\prime$ axis (see Fig.~\ref{fig1}), 
and $a$ is the 
forward-backward asymmetry originating from the parity-violating
coupling to the $Z$ boson.
We recently showed~\cite{peng16} that Eq.~(\ref{eq:eq2}) can
be derived from Eq.~(\ref{eq:eq5}) by noting that 
\begin{equation}
\cos \theta_0 = \cos \theta \cos \theta_1 + \sin \theta \sin \theta_1
\cos (\phi - \phi_1).
\label{eq:eq6}
\end{equation}
Substituting Eq.~(\ref{eq:eq6}) into Eq.~(\ref{eq:eq5}), one obtains
\begin{eqnarray}
\frac{d\sigma}{d\Omega} & \propto & (1+\cos^2\theta)+
\frac{\sin^2\theta_1}{2} (1-3\cos^2\theta)\nonumber \\
& + & (\frac{1}{2} \sin 2\theta_1 \cos \phi_1)
\sin 2\theta \cos\phi \nonumber \\
& + & (\frac{1}{2} \sin^2\theta_1 \cos 2\phi_1)
\sin^2\theta \cos 2\phi \nonumber \\
& + & (a \sin \theta_1 \cos \phi_1) \sin\theta \cos\phi
+ (a \cos \theta_1) \cos\theta \nonumber \\
& + & (\frac{1}{2} \sin^2\theta_1 \sin 2\phi_1) \sin^2\theta \sin 2\phi
\nonumber \\
& + & (\frac{1}{2} \sin 2\theta_1 \sin\phi_1) \sin 2\theta \sin\phi
\nonumber \\
& + & (a \sin\theta_1 \sin\phi_1) \sin\theta \sin\phi.
\label{eq:eq7}
\end{eqnarray}
A comparison between Eq.~(\ref{eq:eq2}) and Eq.~(\ref{eq:eq7}) shows 
a one-to-one correspondence for all angular distribution terms.
Moreover, the angular distribution
coefficients $A_0 - A_7$ can now be expressed in terms of the quantities
$\theta_1, \phi_1$ and $a$ as follows:
\begin{align}
A_0 &=  \langle\sin^2\theta_1\rangle & A_1 &= \frac{1}{2} \langle\sin 2\theta_1\cos \phi_1\rangle \nonumber \\
A_2 &=  \langle\sin^2\theta_1 \cos 2\phi_1\rangle & A_3 &= \langle a \sin \theta_1 \cos \phi_1\rangle \nonumber \\
A_4 &=  \langle a \cos \theta_1\rangle & A_5 &=  \frac{1}{2} \langle\sin^2\theta_1 \sin 2\phi_1\rangle \nonumber \\
A_6 &= \frac{1}{2} \langle\sin 2\theta_1 \sin\phi_1\rangle &
A_7 &=  \langle a \sin\theta_1 \sin\phi_1\rangle.
\label{eq:eq8}
\end{align}
The $\langle \cdot \cdot \cdot \rangle$ in Eq.~(\ref{eq:eq8}) is a
reminder that the measured values of $A_i$ at given values of
$q_T$ and $y$ are averaged over events having different 
values of $\theta_1, \phi_1$ and $a$, in general. 
Equation (8) is a generalization of an earlier work~\cite{teryaev} which
considered the special case of $\phi_1 =0$ and $a=0$. 
\begin{figure}[tb]
\centering
\subfigure[]
{\includegraphics*[width=0.23\textwidth]{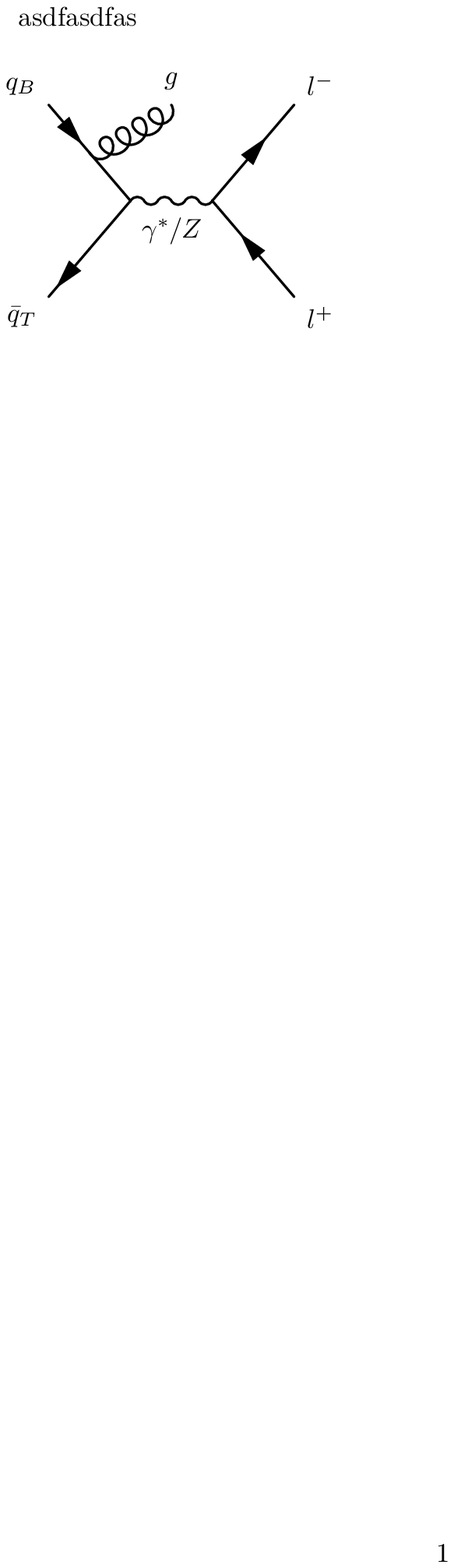}}
\subfigure[]
{\includegraphics*[width=0.23\textwidth]{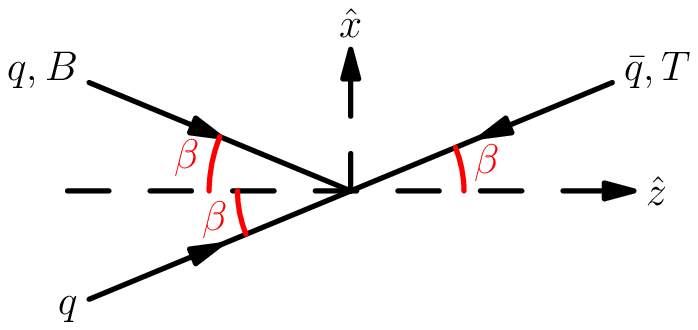}}
\subfigure[]
{\includegraphics*[width=0.23\textwidth]{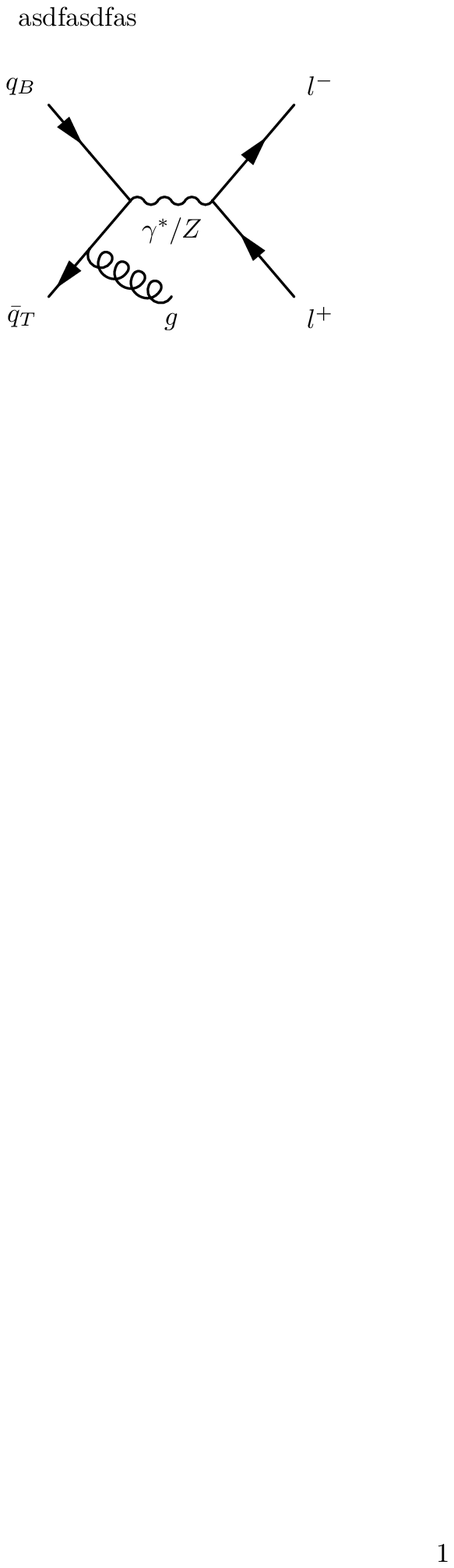}}
\subfigure[]
{\includegraphics*[width=0.23\textwidth]{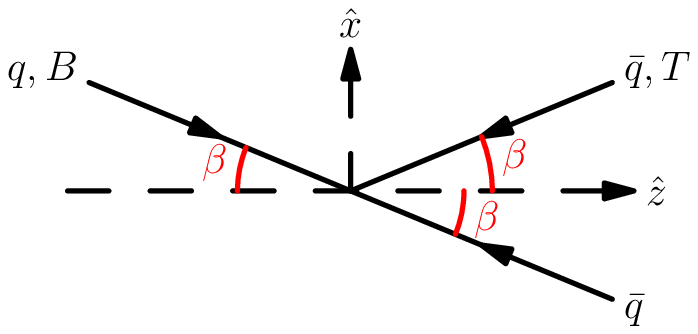}}
\caption{(a) Feynman diagram for $q - \bar q$ annihilation where a gluon
is emitted from a quark in the beam hadron (B). (b) Momentum direction for
$q$ and $\bar q$ in the C-S frame before and after gluon emission. The momentum
direction of $q$ is now collinear with that of $\bar q$. (c) Feynman
diagram for the case where a gluon is emitted from an antiquark in the
target hadron (T). (d) Momentum direction for $q$ and $\bar q$ in the C-S
frame before and after gluon emission for diagram (c).}
\label{fig2}
\end{figure}

The values of $A_0 - A_7$ are bounded by certain limits as a result
of the properties of the trigonometric functions and $|a|<1$.
In particular, we obtain the following relations from Eq.~(\ref{eq:eq8}):
\begin{align}
0 & \le A_0 \le 1 & -1/2 &\le A_1 \le 1/2 \nonumber \\
-1 & \le A_2 \le 1 & -1 &\le A_3 \le 1 \nonumber \\
-1 & \le A_4 \le 1 & -1/2 &\le A_5 \le 1/2 \nonumber \\
-1/2 & \le A_6 \le 1/2 & -1 & \le A_7 \le 1.
\label{eq:eq9}
\end{align}
The bounds on $A_0, A_1, A_2$, together with Eq.~(\ref{eq:eq3}), imply
that 
\begin{equation}
-1/3 \le \lambda \le 1;~~~-1 \le \mu \le 1;~~~-1 \le \nu \le 1.
\label{eq:eq10}
\end{equation}
Some inequality relations among the various coefficients
$A_i$ can also be obtained
from Eq.~(\ref{eq:eq8}). In particular, $A_0$ and $A_2$ satisfy the relation
\begin{equation}
A_0 \ge A_2.
\label{eq:eq11}
\end{equation}
Equation (\ref{eq:eq8}) shows that in the case of $\phi_1 = 0$ or $\pi$, 
i.e., the quark plane and hadron plane
are coplanar, the Lam-Tung relation $A_0 = A_2$ is obtained. When Lam-Tung
relation is violated, $A_0$ must be greater than $A_2$ or,
equivalently, $1-\lambda > 2 \nu$.

While the values of $\theta_1$, $\phi_1$, and $A_i$ depend on the 
specific coordinate system chosen
for the $\gamma^*/Z$ rest frame, it is worth noting that the relations
in Eqs.~(\ref{eq:eq8})-(\ref{eq:eq11}) are independent of this choice, as
long as the $\hat x$ and $\hat z$ axes 
of the reference frame lie within the hadron plane. 
Examples of such reference frames include the Collins-Soper, Gottfried-Jackson,
and the helicity frames.
As a consequence, if the Lam-Tung relation is satisfied
(or violated) in any of these frames, it will be satisfied (or violated) in 
all other frames.

\begin{table}[tbp]   
\caption {Angles $\theta_1$ and $\phi_1$ for four cases of gluon emission
in the $q - \bar q$ annihilation process at order-$\alpha_s$. The signs of
$A_0$ to $A_4$ for the four cases are also listed.}
\label{tab:angles}
\begin{center}
\begin{tabular}{|c|c|c|c|c|c|c|c|c|}
\hline
\hline
Case & Gluon emitted from & $\theta_1$ & $\phi_1$ & $A_0$ & $A_1$ & $A_2$ & $A_3$ & $A_4$ \\
\hline
\hline
1 & Beam quark & $\beta$ & 0 & + & + & + & + & + \\
\hline
2 & Target antiquark & $\beta$ & $\pi$ & + & $-$ & + & $-$ & + \\
\hline
3 & Beam antiquark & $\pi - \beta$ & 0 & + & $-$ & + & + & $-$ \\
\hline
4 & Target quark & $\pi - \beta$ & $\pi$ & + & + & + & $-$ & $-$ \\
\hline
\hline
\end{tabular}
\end{center}
\end{table}

As shown in Eq.~(\ref{eq:eq8}), the $q_T$ and $y$ dependencies of the
angular distribution coefficients, $A_i$, are entirely governed by the
$q_T$ and $y$ dependencies of $\theta_1, \phi_1$ and $a$.
We first consider the quantities $\theta_1$ and $\phi_1$, ignoring
the small intrinsic transverse momentum, $k_T$, of the partons. At the 
leading-order in $\alpha_s$ ($\alpha_s^0$), the quark axis, $\hat z^\prime$, 
is collinear with the $\hat z$ axis. Hence, the result 
$\theta_1 = 0$ (or $\theta_1 = \pi$) is obtained, 
and Eq.~(\ref{eq:eq8})
shows that all $A_i$ except $A_4$ vanish. 

\begin{figure}[tb]
\centering
\subfigure[]
{\includegraphics*[width=0.23\textwidth]{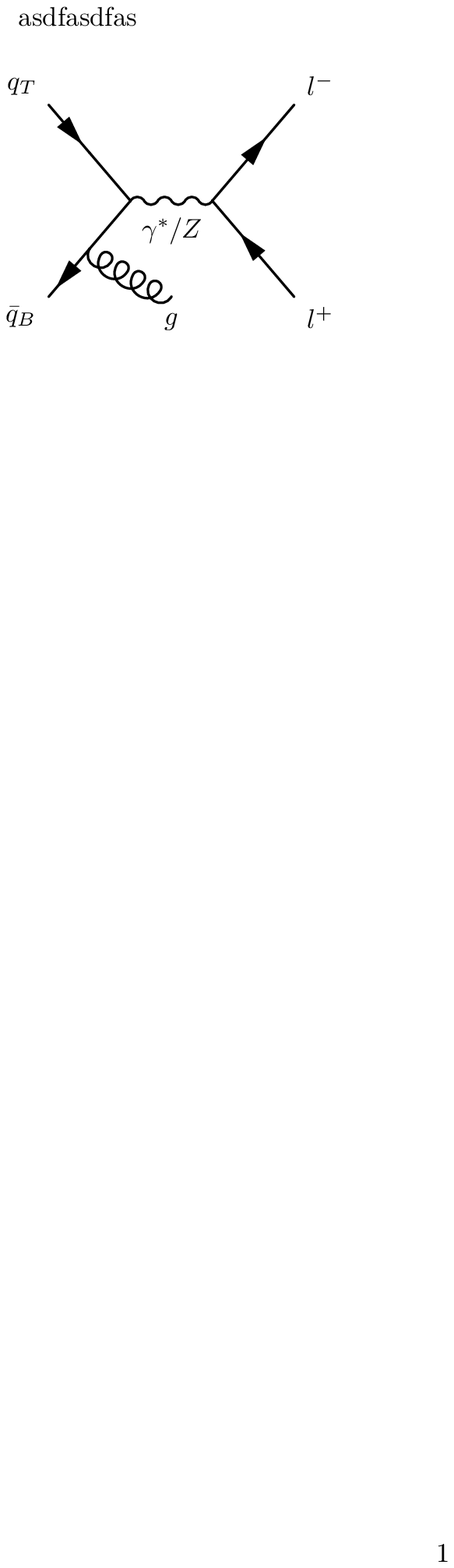}}
\subfigure[]
{\includegraphics*[width=0.23\textwidth]{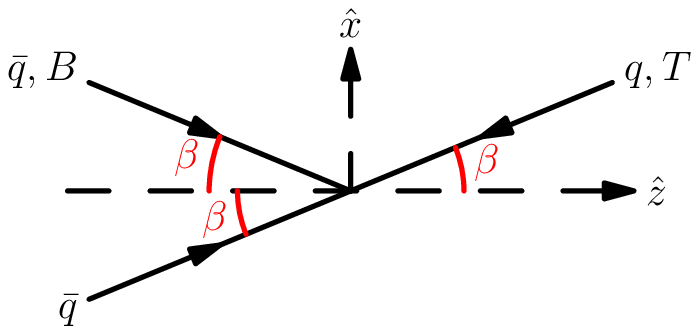}}
\subfigure[]
{\includegraphics*[width=0.23\textwidth]{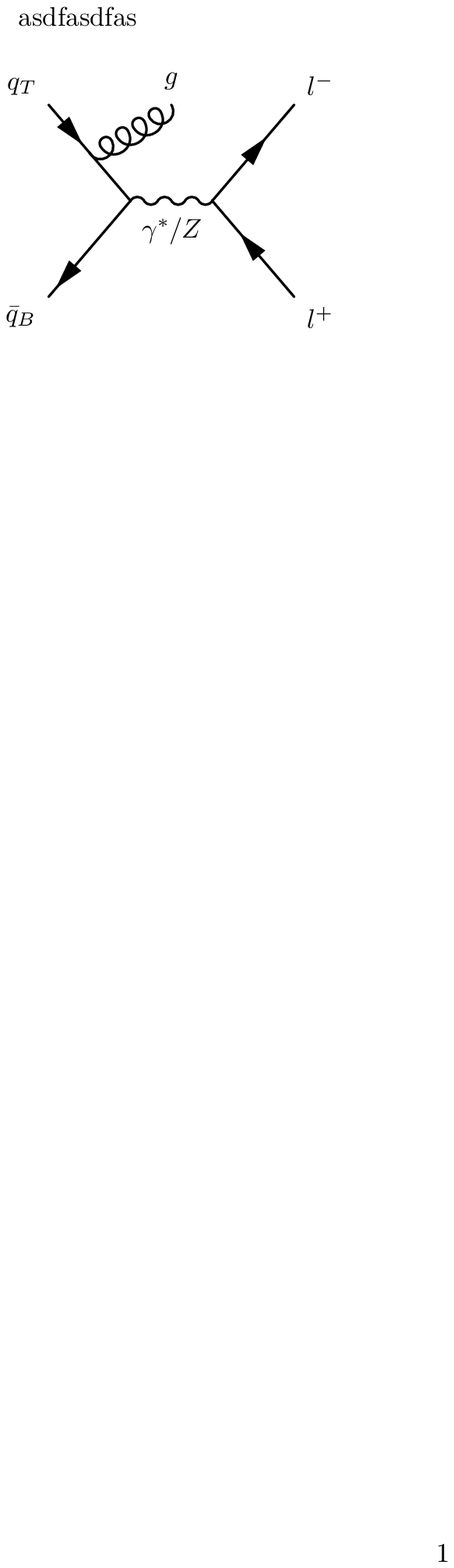}}
\subfigure[]
{\includegraphics*[width=0.23\textwidth]{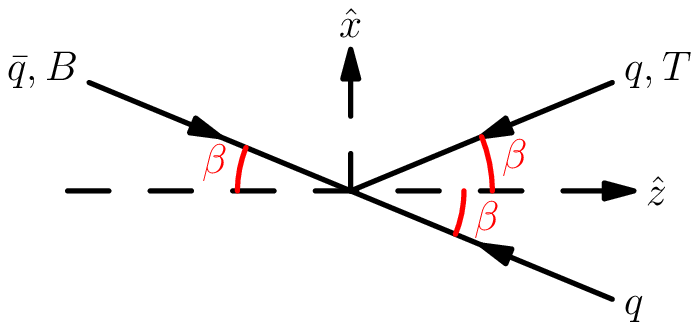}}
\caption{(a) Feynman diagram for $q - \bar q$ annihilation where a gluon
is emitted from an antiquark in the beam hadron (B). (b) Momentum direction for
$q$ and $\bar q$ in the C-S frame before and after gluon emission. The momentum
direction of $q$ is now collinear with that of $\bar q$. (c) Feynman
diagram for the case where a gluon is emitted from a quark in the
target hadron (T). (d) Momentum direction for $q$ and $\bar q$ in the C-S
frame before and after gluon emission for diagram (c).}
\label{fig3}
\end{figure}

At the next-to-leading order (NLO),
$\alpha_s$, a hard gluon or a quark (antiquark) is emitted 
so that $\gamma^*/Z$ acquires nonzero $q_T$. Figure~\ref{fig2}(a) shows 
a diagram for the $q - \bar q$ annihilation process
in which a gluon is emitted
from the quark in the beam hadron. In this case, 
the momentum vector of the quark
is modified such that it becomes opposite to the antiquark's momentum 
vector in the rest frame of $\gamma^*/Z$.
Since the antiquark's momentum direction is the same as the target hadron's 
momentum direction, the $z^\prime$ axis 
is along the direction of $- \vec p_T$ (see Fig.~\ref{fig2}(b)). 
From Fig.~\ref{fig1}, it is evident
that $\theta_1 = \beta$ and $\phi_1 = 0$ in this case. Similarly,
for the case of Fig.~\ref{fig2}(c), where a gluon is emitted from an antiquark
in the target hadron, one obtains $\theta_1 = \beta$ and $\phi_1 = \pi$,
as illustrated in Fig.~\ref{fig2}(d). 
Analogous results with $\theta_1 = \pi - \beta$ and $\phi_1 = 0$ 
(or $\phi = \pi$)
can be found when the roles of beam and target are interchanged,
as illustrated in Fig.~\ref{fig3}.
Table I lists the values of $\theta_1$ and $\phi_1$ for the 
four cases considered above. 
Given $\theta_1 = \beta$ (or $\theta_1 = \pi - \beta$) and 
$\tan \beta = q_T/ Q$ in the Collins-Soper frame, 
we obtain the following
results, relevant for the coefficients $A_i$ in Eq.~(\ref{eq:eq8}), 
for the 
NLO $q - \bar q$ annihilation processes: 
\begin{align}
\sin \theta_1  &=  q_T /(Q^2 + q^2_T)^{1/2} \nonumber \\
\cos \theta_1 &= \pm Q / (Q^2 + q^2_T)^{1/2} \nonumber \\
\sin^2 \theta_1 &= q^2_T / (Q^2 + q^2_T) \nonumber \\
\sin 2 \theta_1 &= \pm 2 q_T Q / (Q^2 + q^2_T),
\label{eq:eq12}
\end{align}
where the $+$ ($-$) sign corresponds to $\theta_1 = \beta$ 
($\theta_1 = \pi - \beta$). Since $\phi_1 = 0$ or $\pi$, one can
see from Table I and Eq.~(\ref{eq:eq8}) that 
the Lam-Tung relation, $A_0 = A_2$,
is satisfied. Moreover, $A_5 - A_7$ must vanish, since they are
proportional to $\sin \phi_1$ or $\sin 2 \phi_1$, which are
identically zero.

\begin{figure}[tb]
\centering
\subfigure[]
{\includegraphics*[width=0.23\textwidth]{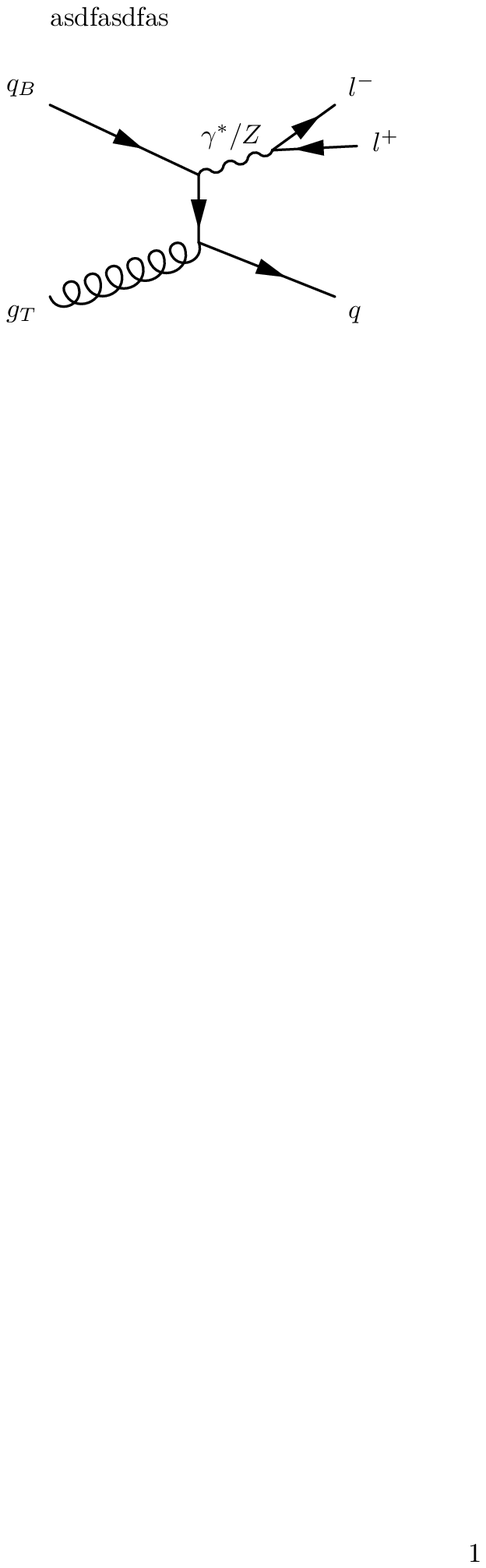}}
\subfigure[]
{\includegraphics*[width=0.23\textwidth]{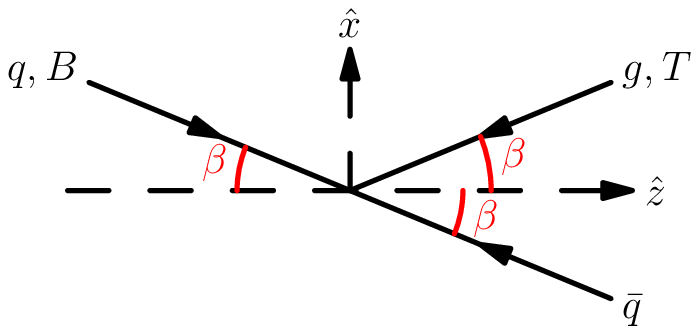}}
\subfigure[]
{\includegraphics*[width=0.23\textwidth]{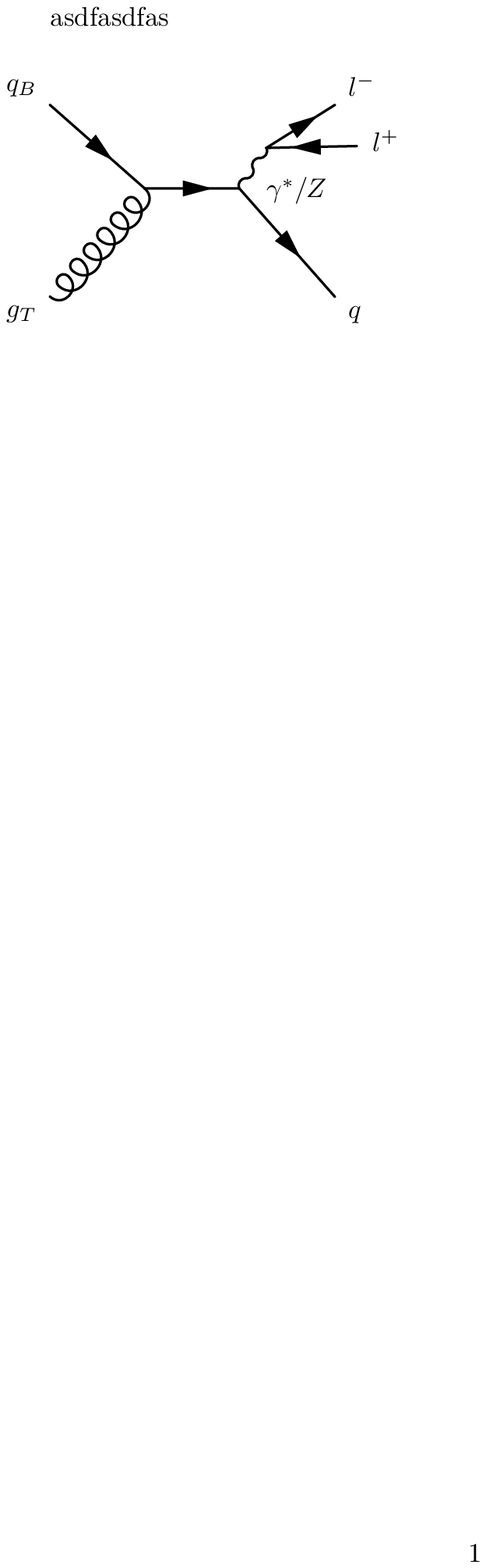}}
\subfigure[]
{\includegraphics*[width=0.23\textwidth]{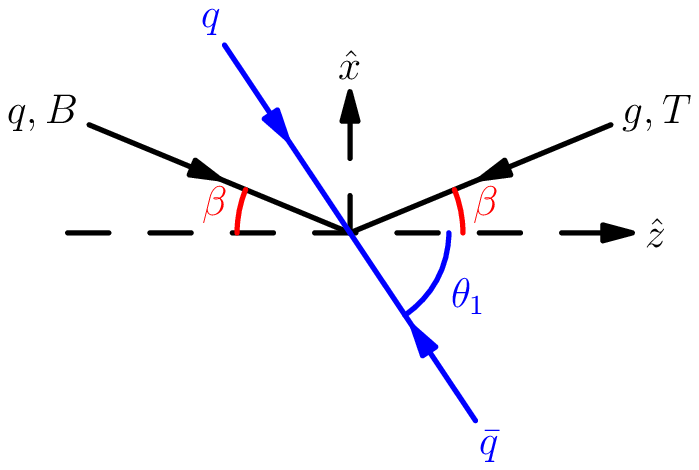}}
\caption{ (a) Feynman diagram for $qG$ Compton process where a quark from the
beam hadron annihilates with an antiquark from the splitting of a gluon
in the target hadron. (b) Momentum direction of $q$, $\bar q$ and gluon in the
C-S frame before and after gluon splitting.
(c) Feynman diagram for $qG$ fusing into a quark which then emits
a $\gamma^*/Z$. (d) Momentum direction of 
$q$, $\bar q$ and gluon before and after the $qG$ fusion.}
\label{fig4}
\end{figure}

We next consider the Compton process at NLO. 
Unlike the cases for the $q - \bar q$ initial state shown in 
Figs. 2 and 3 where a hard gluon is emitted, a hard quark or
antiquark will now accompany the $\gamma^*/Z$ final state.
Fig. 4(a) shows the diagram in which a gluon from the target hadron 
splits into a $q - \bar q$ pair and the quark from the beam hadron annihilates 
with the antiquark into a $\gamma^*/Z$. Since the momentum vector of the 
quark in the beam hadron is unchanged, $\theta_1 = \beta$ and 
$\phi_1 = \pi$, as shown in Fig. 4(b). This result is identical to that
for the $q \bar q$ initial state shown in Fig. 2(d). 
Analogous results are obtained when gluon is emitted from the
beam hadron, or when an antiquark replaces the quark in the
initial state. However, a 
different situation is shown in Fig. 4(c), where the quark and gluon
fuse into a quark, which then emits a $\gamma^*/Z$. As indicated in
Fig. 4(d), $\theta_1$ must satisfy $\beta \le \theta_1 \le \pi - \beta$,
since the momenta of the initial quark and gluon combine vectorially,
resulting in a $\theta_1$ within these limits. Therefore, the two
distinct Compton processes would lead to a mean $\theta_1$ larger than
$\beta$, with the exact value governed by the relative weight of these
two processes. It was shown by Thews~\cite{thews} that, to a very good
approximation, $A_0$ satisfies the relation,
$A_0 = 5 q^2_T / (Q^2 + 5q^2_T)$. Since $A_0 = \sin^2 \theta_1$,
we obtain, for the $qG$ Compton processes at order $\alpha_s$, the
following expressions
\begin{align}
\sin \theta_1 &=  \sqrt{5} q_T /(Q^2 + 5 q^2_T)^{1/2} \nonumber \\
\cos \theta_1 &= \pm Q / (Q^2 + 5 q^2_T)^{1/2} \nonumber \\
\sin^2 \theta_1 &= 5 q_T^2 / (Q^2 + 5 q^2_T) \nonumber \\
\sin 2 \theta_1 &= \pm 2 \sqrt{5}q_T Q / (Q^2 + 5 q^2_T).
\label{eq:eq13}
\end{align}
The $+$ and $-$ sign corresponds to $\theta_1 \le \pi/2$ and $\theta_1 \ge
\pi/2$, respectively. 

\begin{figure}[tb]
\includegraphics*[width=1.05\linewidth]{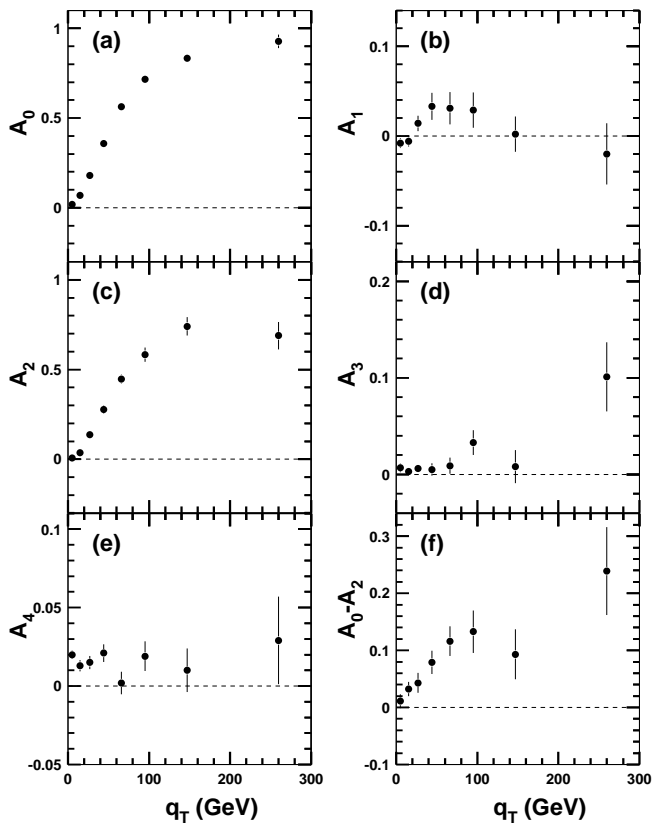}
\caption{The CMS data~\cite{cms} on angular distribution coefficients
$A_i$ versus $q_T$ for $|y| < 1.0$.}
\label{fig5}
\end{figure}

We now consider the parity-violating forward-backward asymmetry,
$a$, in Eqs.~(\ref{eq:eq5}) and (\ref{eq:eq8}). 
The electroweak theory for $Z$ boson production
gives $a = 2 A_f A_{f^\prime}$ for the $f + \bar f \to Z 
\to f^\prime + \bar f^\prime $
process, where $A_f$ is given as
\begin{equation}
A_f = \frac{2 C^f_V C^f_A}{(C^f_V)^2 + (C^f_A)^2}.
\label{eq:eq14}
\end{equation}
The vector $C^f_V$ and axial vector $C^f_A$ couplings 
for $Z$ boson to fermion $f$ are, respectively, 
$I^3_W - 2Q \sin ^2 \theta_W$ and $I^3_W$, where $I^3_W$ and $\theta_W$
denote the weak-isospin third component and the Weinberg angle.
Using $\sin ^2 \theta_W = 0.2315$, then Eq.~(\ref{eq:eq14}) 
gives $a= 0.211$ for
$u \bar u \to Z \to l^- l^+$, and $a=0.299$ for $d \bar d \to Z \to
l^- l^+$, where $l$ refers to $e$ or $\mu$. We note that $a$ has
a positive value. Moreover, depending on the relative
weight between the $u \bar u$ and the $d \bar d$ contributions, one
expects the mean value of $a$ to vary between these two limits.  

\begin{figure}[tb]
\includegraphics*[width=\linewidth]{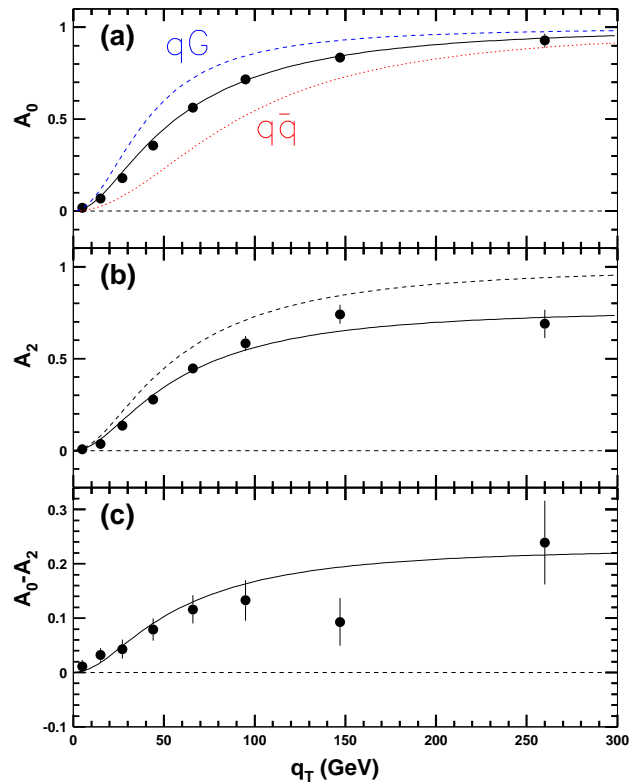}
\caption{Comparison between the CMS data~\cite{cms} on 
$A_0$, $A_2$ and $A_0-A_2$ with calculations. Curves correspond
to calculations described in the text.}
\label{fig6}
\end{figure}

\section {III. Transverse momentum 
dependencies of angular distribution coefficients}

We now compare the $\gamma^*/Z$ production data at the LHC with 
calculations based on the results obtained in Sec. II. 
The LHC data cover a broad range in the dilepton's $q_T$
and rapidity $y$ ($0< q_T < 600$ GeV and $0 < |y| < 3.5$).
For simplicity, we only consider the CMS data in this work.
The ATLAS data contain both the $\mu^- \mu^+$ and
$e^- e^+$ dilepton events, doubling the statistics compared to the
$\mu^- \mu^+$ data sample in CMS. However, the procedure
of ``regularization" adopted by the ATLAS Collaboration 
introduces model dependencies associated with the
theoretical calculations used in the procedure. Although the
tabulated uncertainties of the ATLAS data~\cite{atlas} are significantly
smaller than that of the CMS data~\cite{cms}, it is difficult
to assess the systematic uncertainties associated with the procedure
of ``regularization". We therefore prefer to compare our calculations
with the results of CMS, where a conventional analysis procedure 
without ``regularization" is adopted.

Figure~\ref{fig5} shows the angular distribution coefficients
$A_i$ at the mid-rapidity region $|y| < 1.0$ measured by the 
CMS Collaboration. 
Some salient features in the $q_T$ dependencies of $A_i$ are observed. 
Figure~\ref{fig5} shows that the 
coefficients $A_0 -A_3$ are consistent
with zero at the smallest value of $q_T$.
On the other hand, the coefficient $A_4$ is nonzero at
$q_T \to 0$.
The values of $A_5 - A_7$ are found by the CMS Collaboration
to be consistent with zero~\cite{cms}.
In order to understand these
general features of the angular distribution coefficients,
Eq.~(\ref{eq:eq8}) suggests that one could
examine the properties of the quantities $\theta_1$ and $\phi_1$.

From Eqs.~(\ref{eq:eq8}),~(\ref{eq:eq12}),~(\ref{eq:eq13}),
noting that $\phi_1 = 0$ or $\pi$
and the $\gamma^*/Z$ cross sections are dominated by the NLO 
$q \bar q$ and $q G$ processes depicted in Figs. 4 and 5, one can 
readily predict the following patterns for the
$q_T$ dependencies of $A_0$ up to $A_4$:


1) As $q_T \to 0$, 
Eqs.~(\ref{eq:eq8}), (\ref{eq:eq12}), (\ref{eq:eq13}) show that 
$A_0, A_1, A_2, A_3$ all approach zero, since $\theta_1 \to 0$.
On the other hand, 
$A_4$ is at its maximal value, since it is proportional to
$\cos \theta_1$. As $q_T \to \infty$, $\theta_1$
approaches the value of $\pi /2$, and $A_0, A_2, A_3$ reach their
maximal values, while $A_1$ and $A_4$ approach zero. As shown
in Fig.~\ref{fig5}, the data are consistent with these expectations.

2) According to Eqs.~(\ref{eq:eq8}), (\ref{eq:eq12}), (\ref{eq:eq13})
the values of $A_0$ would go from zero at $q_T =0$ to unity
as $q_T \to \infty$. At all values of $q_T$, one expects
$A_2 \le A_0$. In the case of $\cos 2\phi_1 = 1$, which occurs for
the NLO processes as discussed above, the Lam-Tung relation,
$A_0 = A_2$ is satisfied. When the Lam-Tung relation is violated,
$A_0 \ne A_2$ (or $1-\lambda \ne 2 \nu$), it is expected that only
$A_0 - A_2 > 0$ (or $1-\lambda - 2 \nu > 0$), not the
alternative inequality $A_0 - A_2 < 0$, can occur. These expectations are
consistent with the data shown in Fig.~\ref{fig5}.

3) As $A_1$ is proportional to $\sin 2 \theta_1$, it would first
increase with $q_T$, reaching a maximum, and then decrease. This is
in contrast to $A_0, A_2,$ and $A_4$, which are expected to increase
with $q_T$ monotonically. Similarly, $A_4$ would decrease monotonically
with $q_T$, as it is proportional to $\cos \theta_1$. The data are
consistent with these expected trends.

4) The upper and lower bounds on $A_i$, listed in Eq.~(\ref{eq:eq9}).
are well satisfied by the data.

\noindent We next compare the CMS data on the angular distribution coefficients
$A_0$ to $A_4$ with calculations based on the intuitive geometric
picture discussed above. 


Figure~\ref{fig6}(a) shows the values of $A_0$
versus $q_T$ for $|y| < 1.0$. The dotted and dashed curves 
correspond to calculations
using Eq.~(\ref{eq:eq8}) and Eqs.~(\ref{eq:eq12}), (\ref{eq:eq13}) for 
the $q \bar q$ and $qG$ processes, 
$A_0 = q^2_T/(Q^2 + q^2_T)$ and $A_0 = 5q^2_T / (Q^2 + 5 q^2_T)$,
respectively.
Note that the $q \bar q$ process alone underestimates $A_0$, while the
$qG$ process overestimates it.
Since these two processes contribute incoherently to 
the $\gamma^*/Z$ production
due to their distinct initial and final states (see Figs. 2-4),
the observed $A_0$ is the result of an incoherent sum of these two
processes. A best fit to the 
data, shown as the solid curve in Fig.~\ref{fig6}(a), is obtained with
a mixture of 58.5 $\pm$ 1.6\% $qG$ and 41.5 $\pm$ 1.6\% $q \bar q$ processes.
The excellent agreement between the 
data and the calculation
lends support to the adequacy of this intuitive geometric picture.
It also suggests that higher-order QCD processes
do not affect the values of $\theta_1$ (and $A_0$) significantly.

Figure~\ref{fig6}(b) displays $A_2$ versus $q_T$ for the $|y|<1.0$
data from CMS. Eq.~(\ref{eq:eq8}) shows that the value of $A_2$ 
should be identical
to that of $A_0$ if $\phi_1 = 0$ or $\pi$. The dashed curve in 
Fig.~\ref{fig6}(b) is identical to the solid curve in Fig.~\ref{fig6}(a),
obtained with a mixture of 58.5\% $qG$ and 41.5\% $q \bar q$ processes. 
The deviation of the dashed curve from
the data shows that the Lam-Tung relation, $A_0 = A_2$,
is violated. From Eq.~(\ref{eq:eq8}), it is evident
that this violation is due to $\phi_1 \ne 0$ or $\pi$, namely, the quark
and hadron planes are not coplanar. This noncoplanarity is caused by 
higher-order processes, in which multiple partons accompany the 
$\gamma^*/Z$
in the final state. The hadron plane then contains the vector sum of
multiple partons, and is in general not coplanar with respect to
the quark plane. 
The effect of the noncoplanarity is to reduce the value of $A_2$
with respect to that of $A_0$. The solid curve in Fig. 6(b),
obtained with an overall reduction factor of 0.77, describes the
CMS $A_2$ data well. This reduction factor,
originating from the $\cos 2 \phi_1$ factor, indicates that the 
effective value of the
noncoplanarity angle, $\phi_1$, is around $20^\circ$.
Figure~\ref{fig6}(c) shows the $q_T$ dependence of $A_0 - A_2$ for 
$|y| < 1.0$. The violation of the Lam-Tung relation, reflected by the nonzero
values of $A_0 - A_2$, is well described by the solid curve taking into account
the overall reduction factor of 0.77 for $A_2$.
\begin{figure}[tb]
\includegraphics*[width=\linewidth]{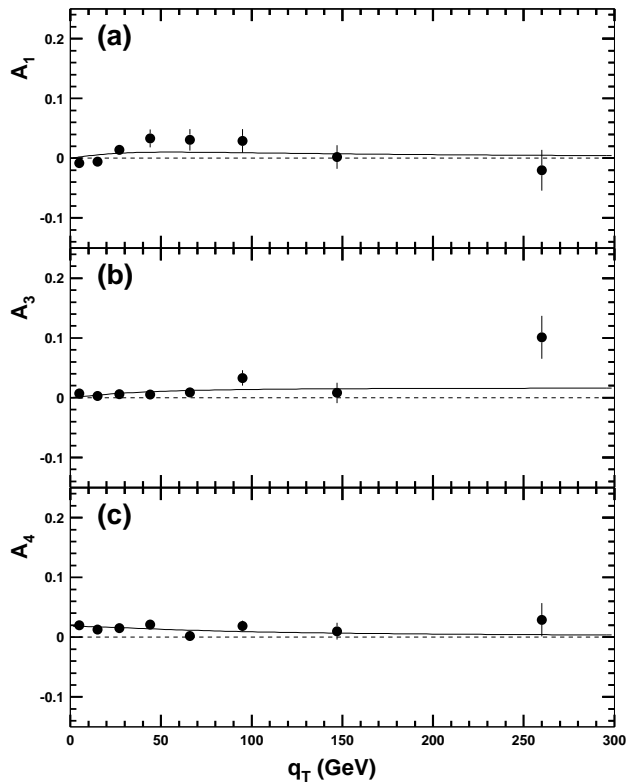}
\caption{Comparison between the CMS data~\cite{cms} on 
$A_1, A_3$ and $A_4$ at $|y| < 1.0$ with calculations. Curves correspond
to calculations described in the text.}
\label{fig7}
\end{figure}

We next consider the coefficient $A_1$. From Eq.~(\ref{eq:eq3}),
the coefficient $A_1$ is related to the parameter $\mu$
measured in fixed-target Drell-Yan experiments. In $p p$ collision,
$A_1$ is odd under $y \leftrightarrow -y$ exchange. Figure 7(a) shows 
the $q_T$ dependence
of $A_1$ measured at CMS. The sign of $A_1$ measured 
at negative $y$ is flipped before combining
it with $A_1$ measured at positive $y$. 
Equation (\ref{eq:eq8}) shows that
$A_1$ is given as 1/2$\langle \sin 2 \theta_1 \cos \phi_1 \rangle$.
The values of $\sin 2 \theta_1$ are given in Eqs.~(\ref{eq:eq12})
and~(\ref{eq:eq13}) for the $q \bar q$ and $qG$ processes, and 
$\phi_1 = 0$ (or $\pi$). For various cases as listed in Table I, one can
calculate the values of $A_1$ for the four cases. Depending
on the value of $\phi_1$, the sign of $A_1$ can be positive or negative,
as shown in Table I. Hence, one expects a significant cancellation
among contributions from processes with $\phi_1 =0$ or $\phi_1 = \pi$.
The solid curve in Fig. 7(a) is obtained with the following expression
\begin{equation}
A_1 = r_1 [f \frac{q_T Q}{Q^2 + q^2_T} + (1-f) \frac{\sqrt{5} q_T Q}
{Q^2 + 5 q^2_T}],
\label{eq:eq15}
\end{equation}
where $f$ is the fraction of $q \bar q$ process, $f = 0.415$, deduced
from the $A_0$ data discussed earlier. The $\sin 2 \theta_1$ values for
the $q \bar q$ and $q G$ processes given in Eqs. (12) and (13) are
weighted by $f$ and $1-f$, respectively. The reduction factor $r_1$
represents the combined effect of the partial cancellation discussed
above and the deviation of $\phi_1$ from $0$ or $\pi$ due to higher-order
QCD.  The best-fit value of $r_1$ using Eq. (15) is $r_1 = 0.0215$. The small
value of $r_1$ indicates the presence of a strong cancellation at small
values of $y$. 

\begin{figure}[tb]
\includegraphics*[width=\linewidth]{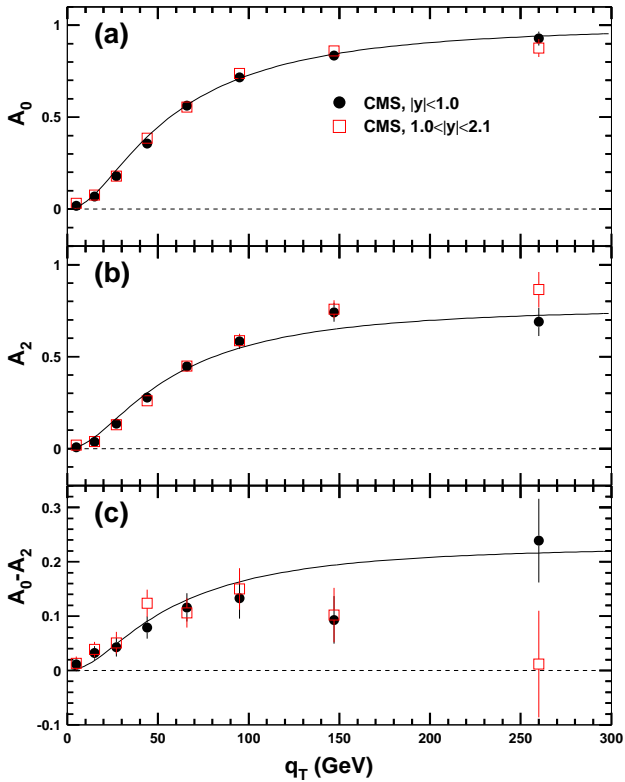}
\caption{Comparison between the CMS data~\cite{cms} on
$A_0$ and $A_2$ at two rapidity regions with calculations. 
Curves correspond
to calculations described in the text.}
\label{fig8}
\end{figure}

Similar considerations also apply to the coefficient $A_3$,
which is also an odd function of $y$ in $p p$ collision.
Both $A_1$ and $A_3$ are sensitive to $\cos \phi_1$. Table I shows
the signs of $A_3$ for four different cases in $q \bar q$ process.
As a parity-violating observable, $A_3$ is also sensitive to the
forward-backward asymmetry parameter $a$. The solid curve in Fig. 7(b)
corresponds to the following expression
\begin{equation}
A_3 = r_3 [f \frac{q_T}{(Q^2 + q^2_T)^{1/2}} + (1-f) \frac{\sqrt{5} q_T}
{(Q^2 + 5 q^2_T)^{1/2}}].
\label{eq:eq16}
\end{equation}
Equation (\ref{eq:eq16}) is analogous to Eq.~(\ref{eq:eq15}), except that the 
reduction factor $r_3$ now includes an additional contribution from $a$. 
The best-fit value, $r_3 = 0.0163$, is obtained.
As shown in Fig. 7(b), the agreement
between the data and this simple calculation is reasonable.

Figure 7(c) shows $A_4$ versus $q_T$ for $|y| < 1.0$. Unlike all other
coefficients, $A_4$ has a nonzero value
as $q_T$ approaches zero.
As discussed earlier, this is well explained by its dependence on 
$\cos \theta_1$, which has a maximal value at $q_T = 0$. The solid curve in
Fig. 7(c) is obtained with the following expression 
\begin{equation}
A_4 = r_4 [f \frac{Q}{(Q^2 + q^2_T)^{1/2}} + (1-f) \frac{Q}
{(Q^2 + 5 q^2_T)^{1/2}}],
\label{eq:eq17}
\end{equation}
where the best-fit value for the reduction factor $r_4$ is 0.0183. 
Both $r_3$ and $r_4$ contain the
parity violating parameter $a$. However, unlike $r_3$, $r_4$ does not
contain the $\cos \phi_1$ term. This qualitatively explains the 
slightly larger
value for $r_4$ than $r_3$. The calculation based on Eq. (17) is in very good 
agreement with the data shown in Fig. 7(c).

\begin{figure}[tb]
\includegraphics*[width=\linewidth]{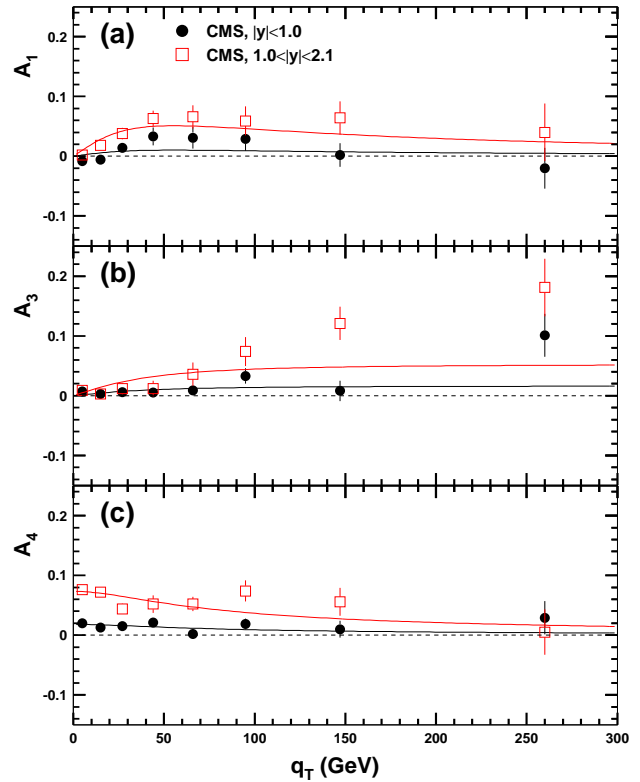}
\caption{Comparison between the CMS data~\cite{cms} on
$A_1, A_3$ and $A_4$ at two rapidity regions with 
calculations. Curves correspond
to calculations described in the text.}
\label{fig9}
\end{figure}






\section {IV. Rapidity dependencies of angular
distribution coefficients}

The CMS Collaboration has reported the rapidity dependencies of $A_i$ for
two bins, $|y| < 1.0$ and
$1.0 < |y| < 2.1$. In this Section, we compare the measured $y$
dependencies with expectations based on our intuitive geometric picture. 
Figure 8
shows that for $A_0$ and $A_2$, there are very weak, if any, rapidity 
dependencies. The solid curves in Fig. 8 are taken from the 
calculations shown in Fig. 6. It is evident that data
at both rapidity bins are well described by a single curve.
The weak rapidity dependence of $A_0$ reflects the fact that $A_0$
only depends on $\theta_1$, which, according to Eqs. (12) and (13),
is independent of the rapidity $y$. However, higher-order QCD effects can
introduce weak rapidity dependence for $A_0$. The weak rapidity 
dependence for $A_2$ shows that $\phi_1$ is weakly $y$ dependent.
Indeed, at order $\alpha_s$, Table I shows that $\cos 2 \phi_1$
is equal to unity for all four cases, independent of the value
of $y$. Again, higher-order QCD will allow $\cos 2 \phi_1$ to
deviate from unity, but the deviation has a very weak $y$ dependence.

\begin{table}[tbp]   
\caption {Reduction factors $r_i$ for $A_1, A_3, A_4$ for
two rapidity bins.}
\label{tab:reduction}
\begin{center}
\begin{tabular}{|c|c|c|}
\hline
\hline
 & $|y| < 1.0$ & $1.0 < |y| < 2.1$ \\
\hline
\hline
$r_1$ & 0.0215 & 0.11 \\
\hline
$r_3$ & 0.0113 & 0.0524 \\
\hline
$r_4$ & 0.0181 & 0.0732 \\
\hline
\hline
\end{tabular}
\end{center}
\end{table}

In striking contrast to $A_0$ and $A_2$, the coefficients $A_1, A_3$ and
$A_4$ exhibit pronounced rapidity dependencies, as shown in Fig. 9. 
A common feature for $A_1, A_3$ and $A_4$ is that they all rise
significantly as $y$ increases. An intuitive explanation for this
strong $y$ dependence is as follows. Table I shows that the various
contributions to $A_1, A_3$ and $A_4$ can be positive or negative,
and each contribution is weighted by the corresponding density distributions
for the interacting partons. At small values of $y$, the momentum fraction
carried by the beam parton, $(x_B)$, is comparable to that of the target
parton, $(x_T)$. Hence the weighting factors for various cases are of
similar magnitude and the net contribution is small due to 
partial cancellations among them. On the other hand, as $y$ becomes large,
$x_B$ becomes significantly larger than $x_T$. Hence, the weighting
factors are now dominated by fewer terms, resulting in less 
cancellation and a larger net result. The various curves shown
in Fig. 9 correspond to calculations using Eqs. (15), (16), (17),
respectively, for $A_1, A_3$ and $A_4$. The CMS data are quite well described
by the best-fit values of $r_1, r_3$, and $r_4$ listed in Table II.


\section {V. Summary and Conclusions}

We have presented an intuitive interpretation for the lepton
angular distribution coefficients for $\gamma^*/Z$ production
measured at the LHC. We first derive the general expression [Eq. (7)] for the
lepton polar and azimuthal angular distributions in the dilepton rest frame,
starting from the azimuthally symmetric lepton angular distribution
[Eq. (5)] with respect to the quark-antiquark axis. We show that the
various angular distribution coefficients are governed by three
quantities, $\theta_1, \phi_1$ and $a$ (Eq. 8). The upper and
lower bounds [Eq. (9)] for the angular distribution coefficients are obtained 
as a result of the expressions in Eq. (8). Similarly, the inequality
relation between $A_0$ and $A_2$, relevant for the violation of the Lam-Tung
relation, is obtained [Eq. (11)].

We then consider the characteristics of the quantities $\theta_1, \phi_1$ 
and $a$. The expressions for $\theta_1$ and $\phi_1$ are obtained
for both the $q \bar q$ and $qG$ processes at order $\alpha_s$.
The $q_T$ dependence of $A_0$ is found to be very well described using the
results for $\theta_1$. It also allows a determination of
the relative fractions of these two processes. This result is noteworthy,
since it shows that a measurement of the angular distribution coefficient
$A_0$ alone could lead to important information on the dynamics of the
production mechanism, namely, the relative contribution of the 
$q \bar q$ annihilation and the $q G$ Compton processes. 

The CMS data clearly show that the Lam-Tung relation, $A_0 = A_2$, is 
violated. The origin of this violation is attributed in our approach to 
the deviation of $\cos 2 \phi_1$ from unity, indicating the noncoplanarity
between the hadron and quark planes. This noncoplanarity is caused by 
higher-order QCD processes. We show that the amount of 
noncoplanarity can be deduced from the $A_0 - A_2$ data directly.
We have also compared our approach with the CMS  
data for other
angular distribution coefficients, $A_1, A_3, A_4$, and found that their
$q_T$ dependencies, governed by the $q_T$ dependence of  
$\theta_1$, can be well described.

We also show that the rapidity dependencies of the $A_i$ can be well
understood in this intuitive approach. In particular, the weak rapidity
dependencies of the $A_0$ and $A_2$, and the pronounced rapidity
dependencies for $A_1, A_3$ and $A_4$ can be explained by the absence
or presence of cancellation effects, which depend strongly on
the rapidity.

We note that the intuitive approach presented in this paper
is by no means a substitute for the perturbative QCD calculations. The
goal of this work is to provide some intuitive explanation of some
salient features present in the lepton angular distribution data. This 
could offer some useful
insights on the origins of many interesting characteristics of the
lepton angular distributions which are being measured at the LHC with
high precision.

The present approach could also be extended to fixed-target Drell-Yan
experiments. Some recent work~\cite{vogelsang} shows the importance of
the perturbative QCD effects even at fixed-target energies. A comparison 
between this intuitive approach and the perturbative QCD calculations
is also of interest. It is also promising to extend this intuitive
approach to some other processes with hadron or lepton beams. 

\section{Acknowledgement}

This work was supported in part by the U.S. National Science Foundation
and the Ministry of Science and
Technology of Taiwan. 





\end{document}